\documentclass{article}
\usepackage{spconf,amsmath,graphicx}
\usepackage{tipa}
\usepackage{xcolor}
\usepackage[hyphens]{url}

\usepackage{pifont}

\usepackage{booktabs}
\usepackage{tabularx}
\usepackage{multirow}
\newcommand{\mytable}{
	\centering
	\renewcommand{\arraystretch}{1.1}
}

\newcolumntype{C}{>{\centering\arraybackslash}X}
\newcolumntype{L}{>{\raggedright\arraybackslash}X}
\newcolumntype{R}{>{\raggedleft\arraybackslash}X}
\newcolumntype{P}[1]{>{\raggedright\arraybackslash}p{#1}}

\newcommand{\PreserveBackslash}[1]{\let\temp=\\#1\let\\=\temp}
\newcolumntype{A}[1]{>{\PreserveBackslash\raggedright}p{#1}}
\newcolumntype{B}[1]{>{\PreserveBackslash\centering}p{#1}}

\usepackage{microtype}

\usepackage{cite}
\let\oldbibliography\thebibliography
\renewcommand{\thebibliography}[1]{\oldbibliography{#1}
                                  \setlength{\itemsep}{-0.2mm}  
                                  \vspace*{-0mm}}

\usepackage{enumitem}

\newif\ifCommentsAuthors

\CommentsAuthorstrue
\ifCommentsAuthors

    \definecolor{myred}{rgb}{.8,.0,.0}
    
    \definecolor{myblue}{rgb}{0,0.4,.8}
    \newcommand{\commentmarc}[1]{\textcolor{myblue}{#1}}
    \definecolor{myblue2}{rgb}{0,0,.5}
    
    \definecolor{mcolor}{rgb}{0,0.5,0.1}
    \newcommand{\commentjulz}[1]{\textcolor{mcolor}{#1}}
    \definecolor{mcolor2}{rgb}{0.6,0.5,0.1}
    
    \definecolor{mygreen}{rgb}{.0,.8,.0}
    
    \definecolor{hermancolor}{HTML}{FF6600}

\else
    \definecolor{myred}{rgb}{.8,.0,.0}
    
    \newcommand{\commentmarc}[1]{}
    \newcommand{\commentjulz}[1]{}
    
\fi

\title{A Comparison of Discrete and Soft Speech Units \\for Improved Voice Conversion}
\name{\begin{tabular}{c}Benjamin van Niekerk$^{1,2}$, Marc-André Carbonneau$^1$, Julian Zaïdi$^1$,\\
Matthew Baas$^2$, Hugo Seuté$^1$, Herman Kamper$^2$\end{tabular}}
\address{$^1$Ubisoft La Forge, Montreal, Canada\\$^2$E\&E Engineering, Stellenbosch University, South Africa}

\begin{document}
%
\maketitle
\begin{abstract}
The goal of voice conversion is to transform source speech into a target voice, keeping the content unchanged.
In this paper, we focus on self-supervised representation learning for voice conversion.
Specifically, we compare discrete and soft speech units as input features. 
We find that discrete representations effectively remove speaker information but discard some linguistic content -- leading to mispronunciations.
As a solution, we propose soft speech units learned by predicting a distribution over the discrete units.
By modeling uncertainty, soft units capture more content information, improving the intelligibility and naturalness of converted speech.\footnote{Audio samples available at {\scriptsize \url{https://ubisoft-laforge.github.io/speech/soft-vc/}}}\footnote{Code available at {\scriptsize \url{https://github.com/bshall/soft-vc}}}
\end{abstract}

\section{Introduction}
\label{sec:intro}

Voice conversion systems transform source speech into a target voice, keeping the content unchanged.
From re-creating young Luke Skywalker in The Mandalorian \cite{mandalorian}, to restoring the voice of an Amyloidosis patient \cite{singer2021respeecher}, voice conversion has applications across entertainment, education and healthcare.

In a typical voice conversion system, the goal is to learn features that capture linguistic content but discard speaker-specific details.
We can then replace the speaker information to synthesize audio in a target voice.
While systems trained on parallel data \cite{toda2007voice, tanaka2019atts2s} or text transcriptions \cite{sun2016phonetic, Huang2020} produce convincing results, they require costly data collection and annotation efforts.
Unsupervised voice conversion addresses this issue by learning without labels or parallel speech \cite{kameoka2018stargan, Qian2019}.
However, there is still a gap in quality and intelligibility between unsupervised and supervised systems \cite{zhao2020voice}.

To bridge this gap, recent work investigates self-supervised representation learning for voice-conversion.
Most of these studies focus on discrete speech units \cite{polyak2021speech, van2020vector, huang2021any}.
The idea is that discretization imposes an information bottleneck separating content from speaker details.
While effective at removing speaker information, discretization also discards some linguistic content -- increasing mispronunciations in the converted speech.
Take the word \textit{fin}, for example. 
Ambiguous frames in the fricative \textipa{/f/} may be assigned to incorrect nearby units, resulting in the mispronunciation \textit{thin}.

To tackle this problem, we propose soft speech units.
Using a fine-tuning procedure similar to \cite{hsu2021hubert}, we train a network to predict a distribution over discrete speech units.
By modeling uncertainty in discrete-unit assignments, we aim to retain more content information and, as a result, correct mispronunciations like  \textit{fin-thin}.
This idea is inspired by soft-assignment in computer vision, which has been shown to improve performance on classification tasks \cite{VanGemert2010}.

Focusing on any-to-one voice conversion (i.e., any source speaker to a single target speaker), we compare discrete and soft speech units across two self-supervised methods: contrastive predictive coding (CPC) \cite{oord2018representation} and hidden-unit BERT (HuBERT) \cite{hsu2021hubert}.
Finally, we evaluate discrete and soft units on a cross-lingual voice conversion task.


Our main contributions are as follows:
\begin{itemize}[leftmargin=10pt,topsep=4pt,partopsep=0pt,parsep=0pt]
\item We propose soft speech units for voice conversion and describe a method to learn them from discrete units.
\item We find that soft units improve intelligibility and naturalness compared to discrete speech units.
\item We show that soft units transfer better to unseen languages in cross-lingual voice conversion.
\end{itemize}

\begin{figure*}[t]
\centering
\includegraphics[width=\textwidth]{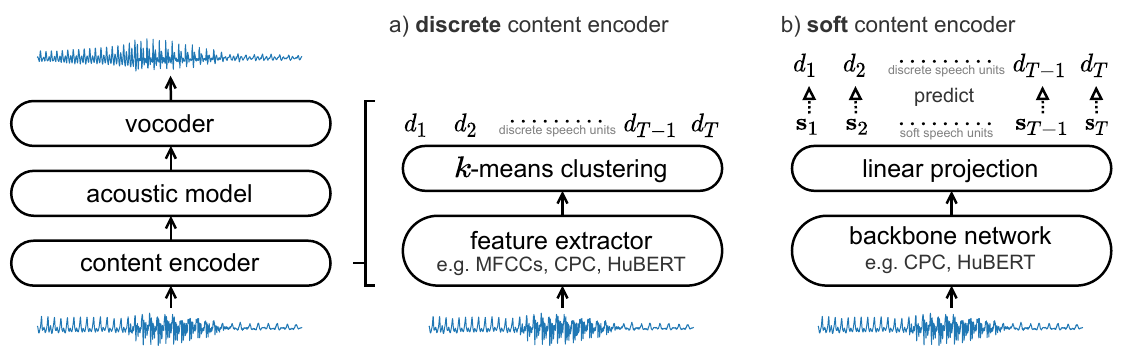}
\vspace*{-12pt}
\caption{
Architecture of the voice conversion system. 
a) The \textbf{discrete} \textit{content encoder} clusters audio features to produce a sequence of discrete speech units.
b) The \textbf{soft} \textit{content encoder} is trained to predict the discrete units.
The \textit{acoustic model} transforms the discrete/soft speech units into a target spectrogram.
The \textit{vocoder} converts the spectrogram into an audio waveform.
}
\vspace*{-5pt}
\label{fig:model}
\end{figure*}

\section{Voice Conversion Systems}
\label{sec:system}

In this section, we describe the voice conversion system we use to compare discrete and soft speech units.
Figure~\ref{fig:model} shows an overview of the architecture.
The system consists of three components: a content encoder, an acoustic model, and a vocoder.
The \textit{content encoder} extracts discrete or soft speech units from input audio (illustrated in Figure~\ref{fig:model}a and \ref{fig:model}b, respectively).
Next, the \textit{acoustic model} translates the speech units into a target spectrogram. 
Finally, the spectrogram is converted into an audio waveform by the \textit{vocoder}.

\subsection{Content Encoders}
\label{sec:content_encoders}



\textbf{Discrete Content Encoder:}
The discrete content encoder consists of feature extraction followed by $k$-means clustering (see Figure~\ref{fig:model}a).
Different feature extractors can be used in the first step -- from low level descriptors such as MFCCs to self-supervised models like CPC or HuBERT.
In the second step, we cluster the features to construct a dictionary of discrete speech units. 
Previous work shows that clustering features from large self-supervised models improves unit quality \cite{hsu2021hubert, nguyen2020zero, vanniekerk2021analyzing} and voice conversion \cite{polyak2021speech}. 
Altogether, the discrete content encoder maps an input utterance to a sequence of discrete speech units $\langle d_1, \ldots, d_T \rangle$.



\vspace{5pt}
\noindent\textbf{Soft Content Encoder:}
For soft speech units, it is tempting to directly use the output of the feature extractor without clustering.
However, previous work \cite{vanniekerk2021analyzing, yang2021superb} shows that these representations contain large amounts of speaker information, rendering them unsuitable for voice conversion (we confirm this in our experiments later).
Instead, we train the soft content encoder to predict a distribution over discrete units.


The idea is that soft speech units provide a middle-ground between raw continuous features and discrete units.
On the one hand, discrete units create an information bottleneck that forces out speaker information.
So to accurately predict the discrete units, the soft content encoder needs to learn a speaker independent representation.
On the other hand, the space of speech sounds is not discrete.
As a result, discretization causes some loss of content information.
By modeling a distribution over discrete units, we aim to keep more of the content information and increase intelligibility.





Figure~\ref{fig:model}b outlines the training procedure for the soft content encoder.
Given an input utterance, we first extract a sequence of discrete speech units $\langle d_1, \ldots, d_T \rangle$ as labels. 
Next, a backbone network (e.g., CPC or HuBERT) processes the utterance.
Then, a linear layer projects the outputs to produce a sequence of soft speech units $\langle \mathbf{s}_1, \ldots, \mathbf{s}_T \rangle$.
Each soft unit parameterizes a distribution over the dictionary of discrete~units:
\[
p(d_t = i \mid \mathbf{s}_t) = \frac{\exp(\text{sim}(\mathbf{s}_t, \mathbf{e}_i) / \tau)}{\sum_{k=1}^K \exp(\text{sim}(\mathbf{s}_t, \mathbf{e}_k) / \tau)},
\]
where $i$ is the cluster index of the $i^\textrm{th}$ discrete unit, $\mathbf{e}_i$ is a corresponding trainable embedding vector, $\text{sim}(\cdot, \cdot)$ computes the cosine similarity between the soft and discrete units, and $\tau$ is a temperature parameter.
Finally, we minimize the average cross-entropy between the distributions and discrete targets $\langle d_1, \ldots, d_T \rangle$ to update the encoder (including the backbone).
At test time, the soft content encoder maps input audio to a sequence of soft speech units $\langle \mathbf{s}_1, \ldots, \mathbf{s}_T \rangle$, which is then passed on to the acoustic model.

\subsection{Acoustic Model and Vocoder}

The acoustic model and vocoder are typical components in a text-to-speech (TTS) system, e.g.,~\cite{Shen2017, zaidi2021}.
For voice conversion, the inputs to the acoustic model are speech units rather than 
graphemes or phonemes. 
The acoustic model translates the speech units (either discrete or soft) into a spectrogram for the target speaker.
Then, the vocoder converts the predicted spectrogram into audio.
There are a range of options for high-fidelity vocoders, including WaveNet \cite{Oord2016} and HiFi-GAN \cite{Kong2020}.


\section{Experimental Setup}

We focus on any-to-one conversion using LJSpeech \cite{ito2018ljspeech} for the target speaker.
We compare discrete and soft speech units across two tasks: intra- and cross-lingual voice conversion.
For intra-lingual conversion, we use the LibriSpeech \cite{panayotov2015librispeech} dev-clean set as source speech.
In the cross-lingual task, we apply systems trained on English to French and Afrikaans data.
For French, we use the CSS10 dataset~\cite{park2019css10}. For Afrikaans, we use data from the South African languages corpus~\cite{van2017rapid}.


\subsection{Model Implementation}

To compare discrete and soft speech units, we implement different versions of the voice conversion system described in Section~\ref{sec:system}. 
For the discrete content encoder, we test CPC and HuBERT as feature extractors.
Similarly, in the soft content encoder, we evaluate CPC and HuBERT as backbones.






CPC learns linguistic representations by predicting future audio segments from a given context.
Based on a contrastive loss, the goal is to distinguish correct future frames from negative examples drawn from other audio files.
We use CPC-big\footnote{\scriptsize \url{https://github.com/facebookresearch/CPC_audio}} \cite{nguyen2020zero} pretrained on the LibriLight unlab-6k set \cite{kahn2020libri}.

HuBERT consists of two steps: acoustic unit discovery followed by masked prediction.
The first step is to construct labels for the
prediction task by clustering either low-level speech features or learned representations from previous training iterations.
The second step is to predict these labels for masked spans of input audio.
We use HuBERT-Base\footnote{\scriptsize \url{https://github.com/pytorch/fairseq}} \cite{hsu2021hubert} pretrained on LibriSpeech-960 \cite{panayotov2015librispeech}.



\vspace{5pt}
\noindent \textbf{Discrete Content Encoder:}
To learn discrete speech units, we apply $k$-means clustering to intermediate representations from CPC-big or HuBERT-base.
We use 100 clusters and estimate their means on a subset of 100 speakers from the LibriSpeech train-clean-100 split.
For CPC-big we cluster the outputs of the second LSTM layer in the context network.
Additionally, we apply the speaker normalization step proposed in \cite{vanniekerk2021analyzing}.
For HuBERT-Base, we use the the seventh transformer layer. We choose these layers because the resulting acoustic units perform well on phone discrimination tests~\cite{hsu2021hubert, nguyen2020zero, vanniekerk2021analyzing}. 


\vspace{5pt}
\noindent \textbf{Soft Content Encoder:}
For the soft content encoder, we use CPC-big and HuBERT-Base as backbones. We fine-tune each model (including the backbone) on LibriSpeech-960 to predict the corresponding discrete speech units.
We train for 25k steps using a learning rate of $2\times10^{-5}$.


\vspace{5pt}
\noindent \textbf{Acoustic Model and Vocoder:}
The acoustic model maps speech units (discrete or soft) to a target spectrogram.
The structure of the model is based on Tacotron 2 \cite{Shen2017} and consists of an encoder and autoregressive decoder.
The encoder is built from a pre-net (two linear layers with dropout), followed by a stack of three 1D-convolutional layers with instance normalization.
For discrete units, we use an initial embedding table to map cluster indexes to vectors.
The decoder predicts each spectrogram frame from the outputs of the encoder and past frames.
We apply a pre-net to the previously predicted frame and concatenate the result with the output of the encoder.
Three LSTM layers with residual connections are then used to model long-term dependencies.
Finally, a linear layer predicts the next spectrogram~frame. 

The vocoder turns the spectrogram frames into an audio waveform.
We choose HiFi-GAN as the vocoder since several papers show that it produces high-quality speech \cite{polyak2021speech, zaidi2021}.

The acoustic model and vocoder are trained on LJSpeech.
We 
downsample the dataset to 16~kHz and 
extract 128-band mel-spectrograms at a hop-length of 10~ms with a Hann window of 64~ms.
We train the acoustic model for 50k steps and select the checkpoint with the lowest validation loss.
For HiFi-GAN, we train using ground-truth spectrograms for 1M steps and then fine-tune on predicted spectrograms for 
500k steps.




\subsection{Baseline Models}

We also compare our voice conversion systems against two common baselines: AutoVC\footnote{\scriptsize \url{https://github.com/auspicious3000/autovc}} \cite{Qian2019} and the Cascaded ASR-TTS\footnote{\scriptsize \url{https://github.com/espnet/espnet/tree/master/egs/vcc20}} system \cite{Huang2020} from the Voice Conversion Challenge 2020~\cite{zhao2020voice}.
AutoVC is an any-to-any voice conversion system based on an autoencoder with a down-sampling bottleneck.
For the Cascaded ASR-TTS model, input speech is first transcribed using an automatic speech recognition (ASR) system. 
The transcripts are then piped to a text-to-speech (TTS) system which generates audio in the target voice.
We fine-tune the released Cascade ASR-TTS checkpoint on LJSpeech.

\subsection{Evaluation Metrics}

To assess the \textit{intelligibility} of the converted speech, we measure word error rate (WER) and phoneme error rate (PER) using an automatic speech recognition (ASR) system.
We use the HuBERT-Large ASR\footnote{\scriptsize \url{https://github.com/huggingface/transformers}} model \cite{hsu2021hubert} for orthographic transcriptions, and Allosaurus~\cite{li2020universal} for phonemic transcriptions.
We convert the ground-truth orthographic transcriptions to phonemes using epitran \cite{mortensen2018epitran}.
Lower error rates correspond to more intelligible speech since it shows that the original words are still recognizable after conversion.
We use Google Cloud Speech-to-Text for Afrikaans ASR and Wav2vec2 for French.

We measure \textit{speaker-similarity} using a trained speaker-verification system.
Given a query and enrollment utterance, the verification system assigns a similarity score indicating whether the speakers match or not.
We use cosine similarity between x-vectors \cite{snyder2018xvectors} as the score.
For evaluation, each converted example is paired
with 50 different enrollment utterances sampled from the target speaker. 
Then we add an equal number of authentic target speaker pairs.
We report equal-error rate (EER), which approaches 50\% when the verification system cannot distinguish between converted and genuine target-speaker utterances (indicating high speaker similarity).



\begin{figure*}[!t]
\centering
\includegraphics[width=\textwidth]{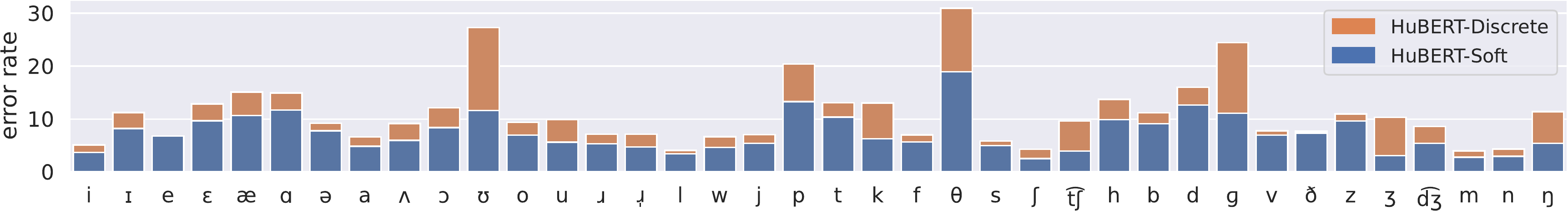}
\vspace*{-12pt}
\caption{Breakdown of PER (\%) per phoneme for HuBERT-Discrete and HuBERT-Soft. 
}
\label{fig:errors}
\vspace*{-5pt}
\end{figure*}

For \textit{naturalness}, we conduct a subjective evaluation based on mean opinion scores (MOS).
We randomly select 20 source speakers from the LibriSpeech dev-clean set and convert 3 utterances per speaker, each between 4 and 15 seconds long.
Evaluators rate the naturalness of the utterances on a five-point scale (from 1=bad to 5=excellent). 
Each sample is judged by at least 4 self-reported English speakers using the CAQE Toolkit \cite{CAQE}.
We report MOS and 95\% confidence intervals.

\section{Experimental Results}



We report results for the English intra-lingual experiments followed by the French and Afrikaans cross-lingual results.

\begin{table}[!b]
	\mytable
	\begin{tabularx}{1.0\linewidth}{@{}lCCCc@{}}
		\toprule
		Method & PER & WER & EER & MOS \\
		\midrule
		HuBERT-Discrete & 10.4 & \hphantom{0}5.4 & 49.8 & 3.69 $\pm$ 0.13 \\
		HuBERT-Soft & \hphantom{0}\textbf{7.8} & \hphantom{0}\textbf{2.6} & 45.6 & \textbf{4.15 $\pm$ 0.12} \\
		HuBERT-Raw-Features & 11.3 & \hphantom{0}2.8 & 27.8 & - \\
		\midrule
		CPC-Discrete & 14.5 & \hphantom{0}8.1 & \textbf{50.0} & 3.37 $\pm$ 0.13 \\
		CPC-Soft & 11.4 & \hphantom{0}3.7 & 41.3 & 3.91 $\pm$ 0.12 \\
		CPC-Raw-Features & 14.6 & \hphantom{0}3.6 & \hphantom{0}5.2 & - \\
		\midrule
		AutoVC \cite{Qian2019} & 58.3 & 73.3 & 13.3 & 1.09 $\pm$ 0.04 \\
		Cascaded ASR-TTS \cite{Huang2020} & \hphantom{0}8.4 & \hphantom{0}7.4 & 46.8 & 3.15 $\pm$ 0.11 \\
		Ground Truth & \hphantom{0}7.9 & \hphantom{0}2.0 & - & 4.57$ \pm$ 0.10 \\
		\bottomrule
	\end{tabularx}
	\vspace*{-10pt}
	\caption{
	Voice conversion results on English.
	PER (\%) and WER (\%) assess intelligibility.
	Speaker-similarity is reported as the EER (\%) of a speaker-verification system.
	MOS from subjective tests indicate~naturalness.
	}
	\label{tbl:evaluation1}
\end{table}

\begin{table}[!b]
	\mytable
	\begin{tabularx}{1.0\linewidth}{@{}lCCCC@{}}
		\toprule
		& \multicolumn{2}{c}{French} & \multicolumn{2}{c}{Afrikaans} \\
        \cmidrule(l){2-3} \cmidrule(l){4-5}
		Method & WER & EER & WER & EER \\
		\midrule
		HuBERT-Discrete & 64.6 & 49.6 & 24.7 & 37.6 \\
		HuBERT-Soft & 28.2 & 33.9 & 12.9 & 28.2 \\
		Ground Truth & 14.6 & - & \hphantom{0}8.0 & - \\
		\bottomrule
	\end{tabularx}
	\vspace*{-10pt}
	\caption{
	Intelligibility (WER) and speaker similarity (EER) results (\%) for the cross-lingual experiments. 
	}
	\label{tbl:evaluation2}
\end{table}


\vspace{5pt}
\noindent \textbf{Intelligibility:}
Table 1 reports intelligibility in terms of PER and WER, with lower rates indicating more intelligible speech.
Compared to discrete units, using soft speech units substantially improves WER (by around 50\% relative) and PER (by over 20\%).
Both Hubert- and CPC-Soft outperform the cascaded ASR-TTS baseline, reaching error rates close to the ground-truth recordings.

Figure~\ref{fig:errors} compares the PER between soft and discrete units in more detail. 
We align phonemic transcriptions of converted speech with the ground-truth and compute an error rate for each phoneme.
Apart from the vowel \textipa{/U/} (c\textbf{oo}k), improvements are primarily in the consonants.
In particular, we see marked improvement in the affricative \textipa{/ \textteshlig/} (\textbf{ch}in), the fricative \textipa{/Z/} (\textbf{j}oke), and the velar stops \textipa{/k/} (\textbf{k}id) and \textipa{/g/}~(\textbf{g}o).

To summarize, the results show that soft assignments capture more linguistic content, improving intelligibility compared to discrete units.


\vspace{5pt}
\noindent \textbf{Speaker Similarity:}
The third column of Table~\ref{tbl:evaluation2} shows speaker similarity results.
An EER of 50\% demonstrates that the speaker verification system cannot differentiate between genuine and converted utterances (i.e., high speaker similarity).
Discrete units obtain near-perfect scores, verifying that they effectively discard source-speaker information.
In comparison, soft units cause a small decrease in similarity.
However, raw features are notably worse, confirming that soft units are a good middle-ground between discrete and continuous features.

\vspace{5pt}
\noindent \textbf{Naturalness:}
The last column of Table~\ref{tbl:evaluation2} reports MOS for naturalness. Across both the CPC- and HuBERT-based models, soft units significantly improve over discrete speech units. 
Hubert-Soft performs best in the listening test, with naturalness scores approaching the ground-truth recordings.
We speculate that better intelligibility explains part of the improvement. 
However, we also think that additional information encoded in the soft units results in more natural prosody.
This could explain why the Cascaded ASR-TTS system scores lower on naturalness despite a better PER and WER than CPC-Discrete.

\vspace{5pt}
\noindent \textbf{Cross-Lingual Voice Conversion:}
Table~\ref{tbl:evaluation2} reports intelligibility and speaker-similarity scores for the cross-lingual task.
We take the best discrete and soft systems from the intra-lingual setting above, and apply them directly to French and Afrikaans test data.
Comparing WER results, we see that soft speech units transfer better to unseen languages. 
However, in the cross-lingual setting, soft units lead to a larger drop in speaker similarity (lower EER). 
We suspect this is because soft units retain more accent information from the source speech.

\section{Conclusion}



We proposed soft speech units to improve unsupervised voice conversion.
We showed that soft units are a good middle-ground between discrete and continuous features -- they accurately represent linguistic content while still discarding speaker information.
In both objective and subjective evaluations, we found that soft units improve intelligibility and naturalness.
Future work will investigate soft speech units for any-to-any voice conversion.

\vfill\pagebreak


\bibliographystyle{IEEEbib}
\bibliography{references}

\end{document}